# Tweets Miner for Stock Market Analysis

Bohdan Pavlyshenko

Electronics department, *Ivan Franko Lviv National University,Ukraine,* Drahomanov Str. 50, Lviv, 79005, Ukraine, *e-mail: b.pavlyshenko@gmail.com*

In this paper, we present a software package for the data mining of Twitter microblogs for the purpose of using them for the stock market analysis. The package is written in R langauge using apropriate R packages. The model of tweets has been considered. We have also compared stock market charts with frequent sets of keywords in Twitter microblogs messages.

## Introduction

The system of microblogs Twitter is one of popular means of interaction among users via short messages (up to 140 characters). Twitter messages are characterized by high density of contextually meaningful keywords. This feature conditions the availability of the study of microblogs by using data mining in order to detect semantic relationships between the main concepts and discussion subjects in microblogs. Very promising is the analysis of predictive ability of time dependences of key quantitative characteristics of thematic concepts in the messages of twitter microblogs. The peculiarities of social networks and users' behavior are researched in many studies. In [1] the microblogging phenomena were investigated. Users' influence in Twitter was studied in [2]. Users' behavior in social networks is analyzed in [3]. In [4] the methods of opinion mining of Twitter corpus were analyzed. Several papers are devoted to the analysis of possible forecasting of events by analyzing messages in microblogs. In [5] it was studied whether public mood, as measured from large-scale collection of tweets, posted on twitter.com, is correlated or predictive for stock markets.

In this paper, we construct a set-theoretic model of key tags of twitter messages. Then we consider our developed software package for data mining of twitter's microblogs. Then we compare the stock market charts with the frequent sets of keywords in Twitter microblogs messages. We analyzed the graphs for frequent itemsets and association rules for specified keywords. We also conducted the Granger test to determine the causation between the time dynamics of frequent sets and the stock price.

## Theoretical Model

Let us consider a model that describes microblogs messages. We have chosen a set of keywords which specifiy the themes of messages and are present in all messages, e.g,

$$kw \in Keywords,\ \ Keywords = \{stock, market\} \qquad (1)$$

Then we define a set of microblogs messages for the analysis:

$$TW^{kw} = \left\{ tw^{(kw)}{}_i \ \ | kw_j \in tw_i,\ \ kw_j \in Keywords \right\} \qquad (2)$$

Our next step is to consider the basic elements of the theory of frequent sets. Each tweet will be considered as a basket of key terms

$$tw_i = \{w_{ij}^{tw}\}. \qquad (3)$$

Such a set is called a transaction. We label some set of terms as

$$F = \{w_j\}. \tag{4}$$

The set of tweets, which includes the set $F$ looks like

$$TW_F^{kw} = \{tw_r \mid F \in tw_r; r = 1,\ldots m\}. \tag{5}$$

The ratio of the number of transactions, which include the set $F$, to the total number of transactions is called a support of $F$ basket and it is marked as $Supp(F)$:

$$Supp(F) = \frac{|TW_F^{kw}|}{|TW^{tw}|}. \tag{6}$$

A set is called a frequent itemset, if its support value is more than the minimum support that is specified by a user

$$Supp(F) > Supp_{\min}. \tag{7}$$

Given the condition (7), we find a set of frequent itemsets

$$L = \{ F_j \mid Supp(F_j) > Supp_{\min} \}. \tag{8}$$

Based on frequent itemsets, we can build association rules, which are considered as

$$X \rightarrow Y, \tag{9}$$

where $X$ is called *antecedent* and $Y$ is called *consequent*. The objects of *antecedent* and *consequent* are the subsets of the frequent set $F$ of considered keywords

$$X \cup Y = F. \tag{10}$$

## Software package for tweets mining

Let us consider the software package for tweets mining, we developed it using R language and appropriate R packages. R GUI environment with working software is presented in the Fig.1. This software allows users to load tweets from a specified list of twitter users, such as: CNN, WSJ, Reuters, Bloomberg, etc., and perform data mining analysis, comparing results with the stock market chart. This package is described and can be loaded in our analytics blog at http://bpavlyshenko.blogspot.com .

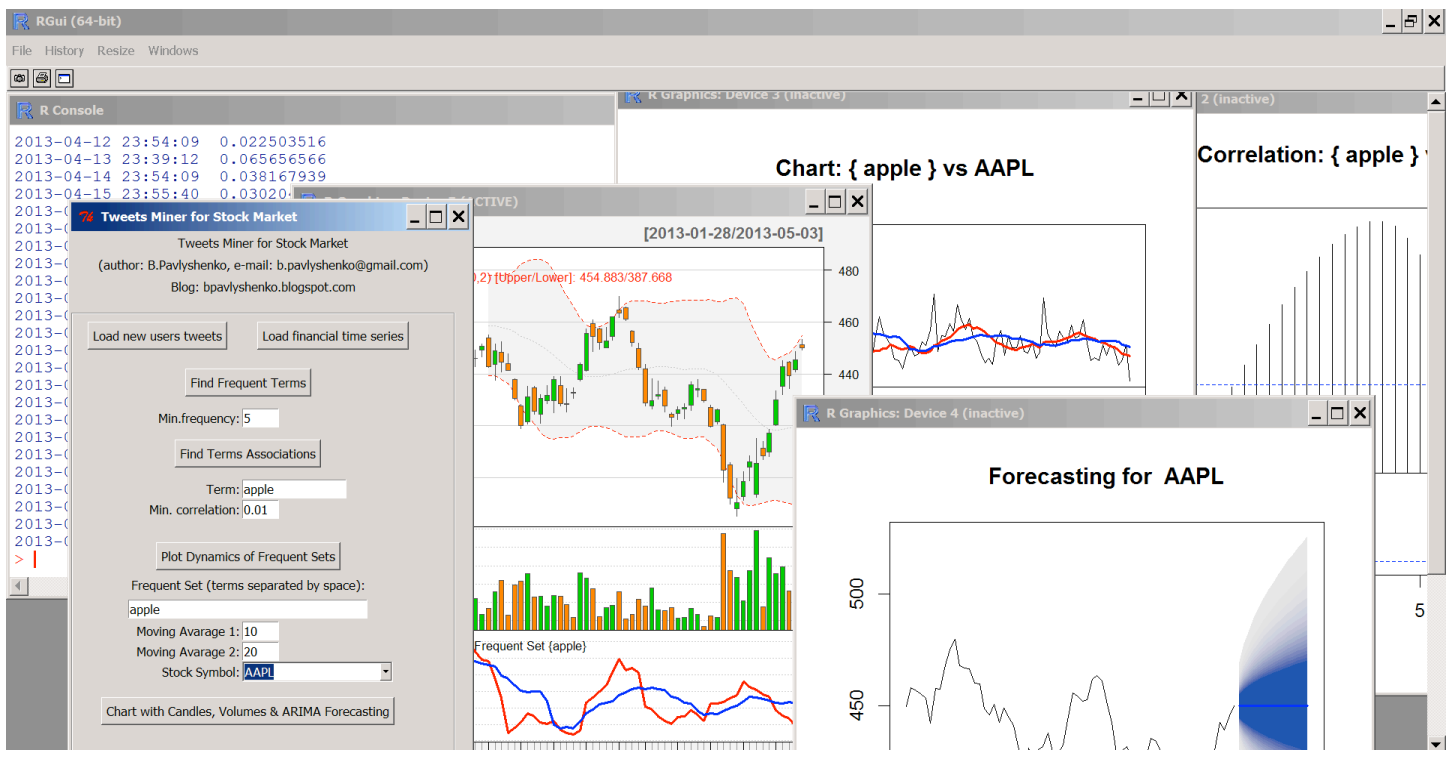

Fig.1 Tweets Miner in the R GUI

The user's interface is followed in the Fig.2. To form frequent sets of keywords for the analysis, one can compose a list of some specific frequent terms, and using this list, one can find the terms which are associated with these frequent terms. E.g. a term "apple" is associated with "aapl", "ipad", "iphone". Then the following frequent sets can be investigated: "apple", "apple aapl", "apple aapl ipad", etc. Before the analysis, one should click on the buttons "Load new users' tweets" and "Load financial time series". Each time the latest tweets only will be loaded, previous tweets are saved in the file and can be used for the next analysis.

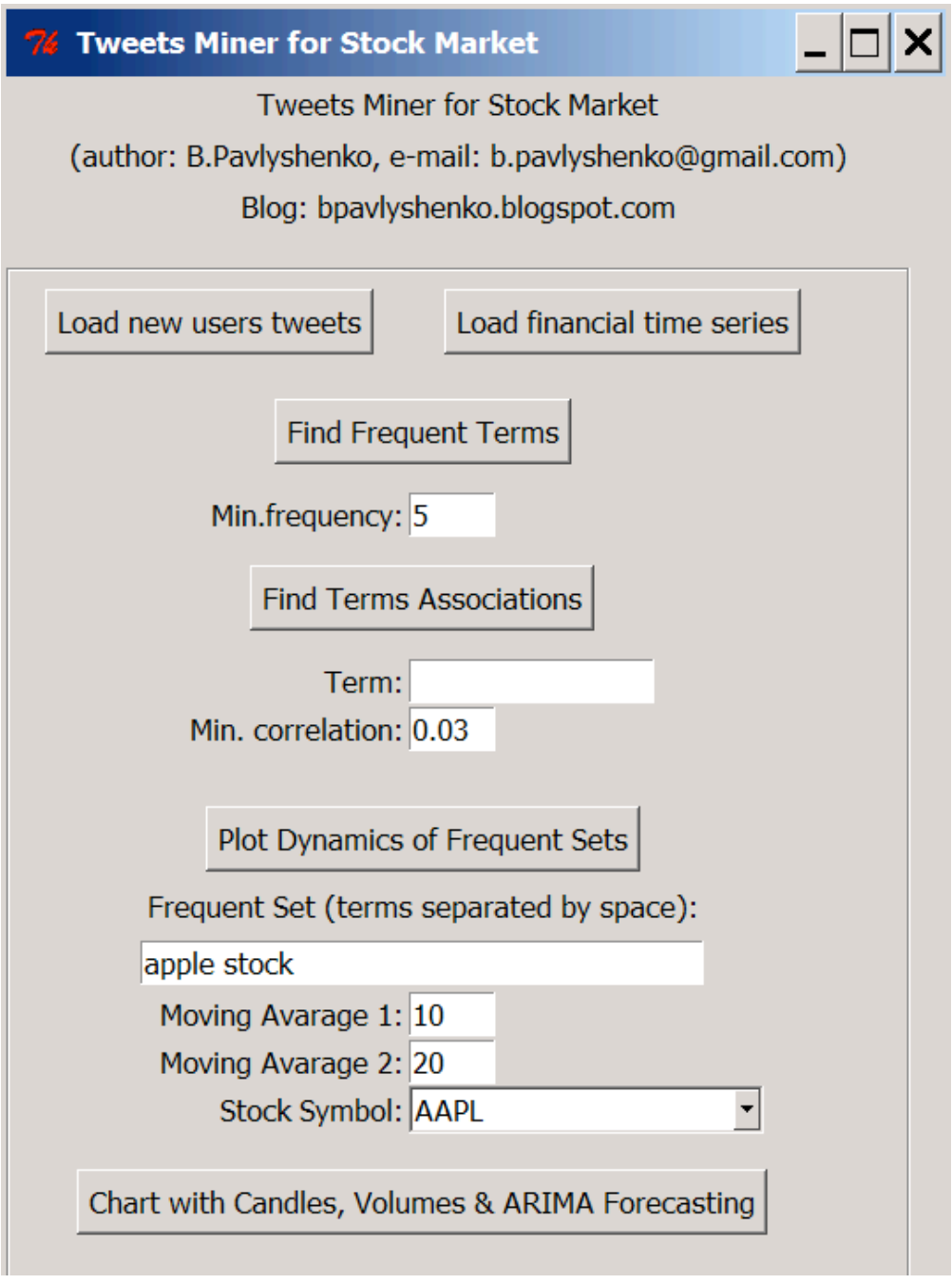

Fig.2 User's interface for Tweets Miner

The analysis can also be performed without loading new tweets and financial time series. In this case the analysis will be carried out for previously loaded tweets and financial time series, which are saved in the data files. When one tries to find frequent terms or associations for the first time per session or after loading new tweets, the program requires several minutes for documents-terms matrix creation. It happens only once per session. A user can browse frequent keywords using a keywords cloud for visualization (Fig.3), and find the associations in the tweets with specified keywords. (Fig.4). To plot the time dynamics of frequent sets of keywords, a users need to specify a frequent set in the lower case, e.g. "apple", "apple aapl", "apple iphone", etc.; then users can choose a stock symbol for comparing time series and choose the time windows for two moving averages, then click on the button "Plot Dynamics of Frequent Sets". The time dynamics of frequent sets of keywords includes two moving averages, which can be used for the trading strategy with the intersection of two moving averages (Fig.5).

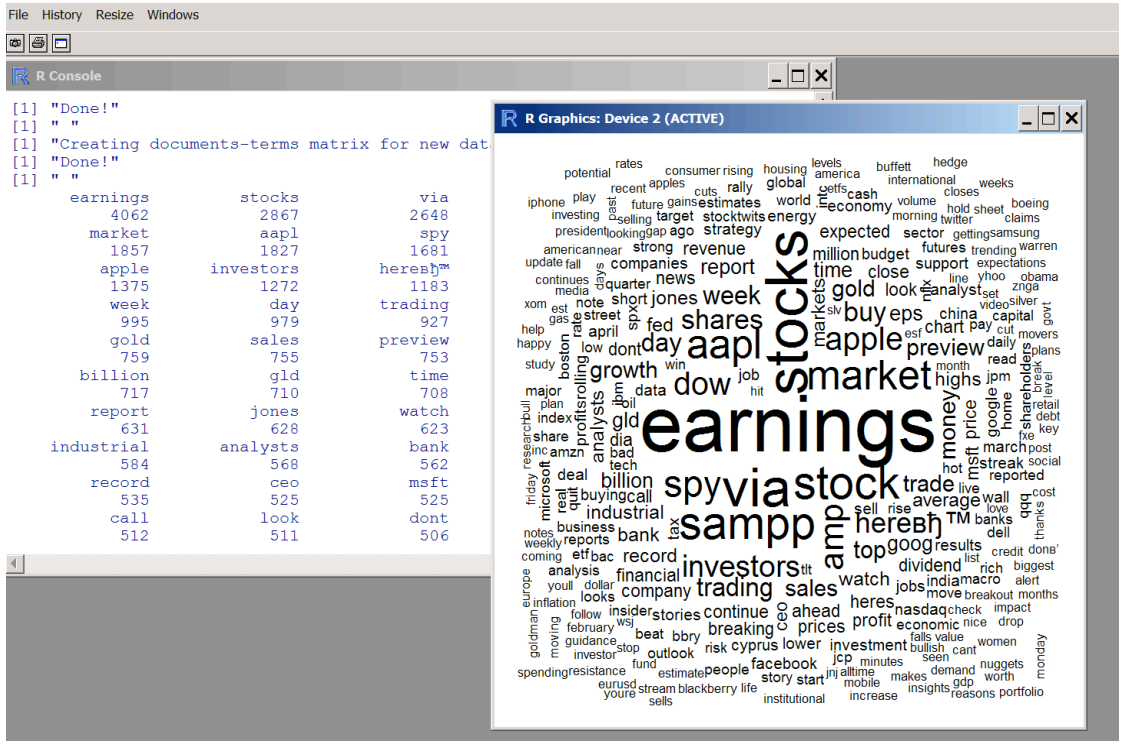

Fig.3 The search of frequent terms list and the terms cloud.

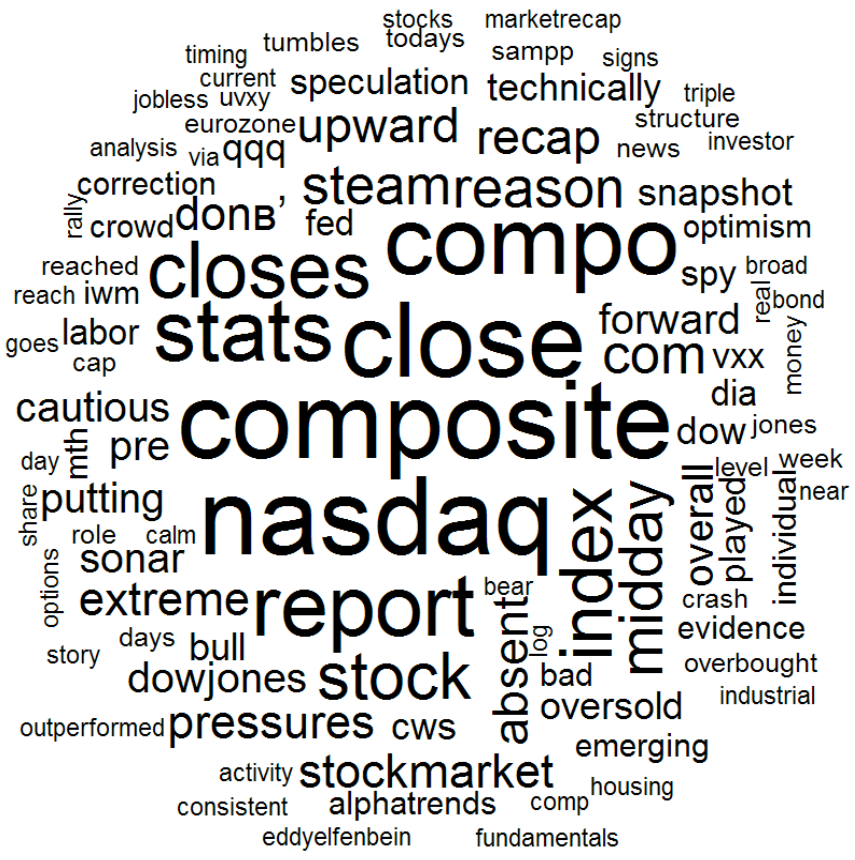

Fig.4 The cloud of keywords, which are associated with
The keyword “market" in the loaded tweets .

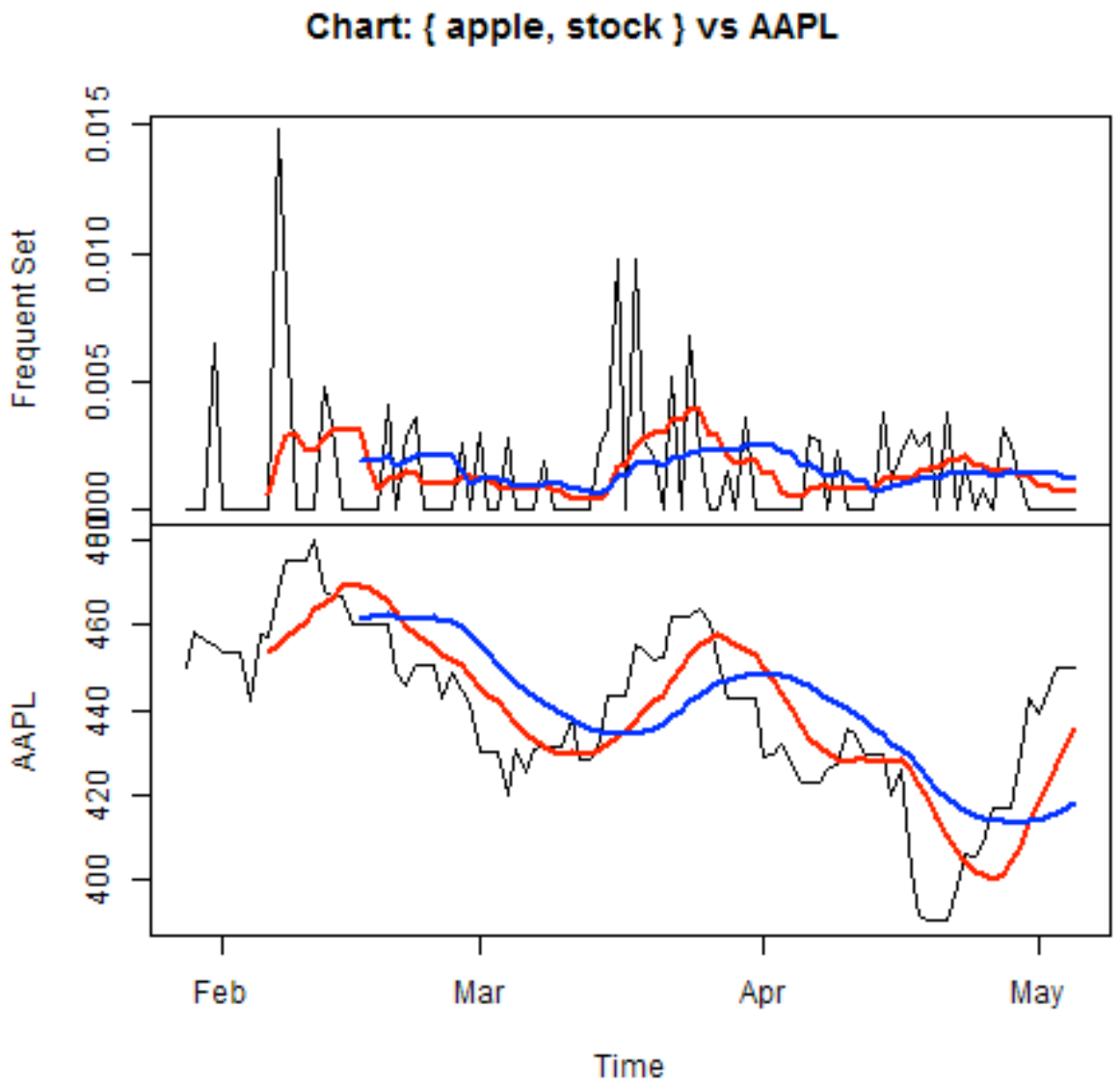

Fig.5 The time dynamics of frequent sets of keywords {apple,stock} and the stock chart for AAPL.

To compare the time dynamics of frequent sets of keywords, they can be plotted as a standard stocks chart with candles and volumes (Fig.6).

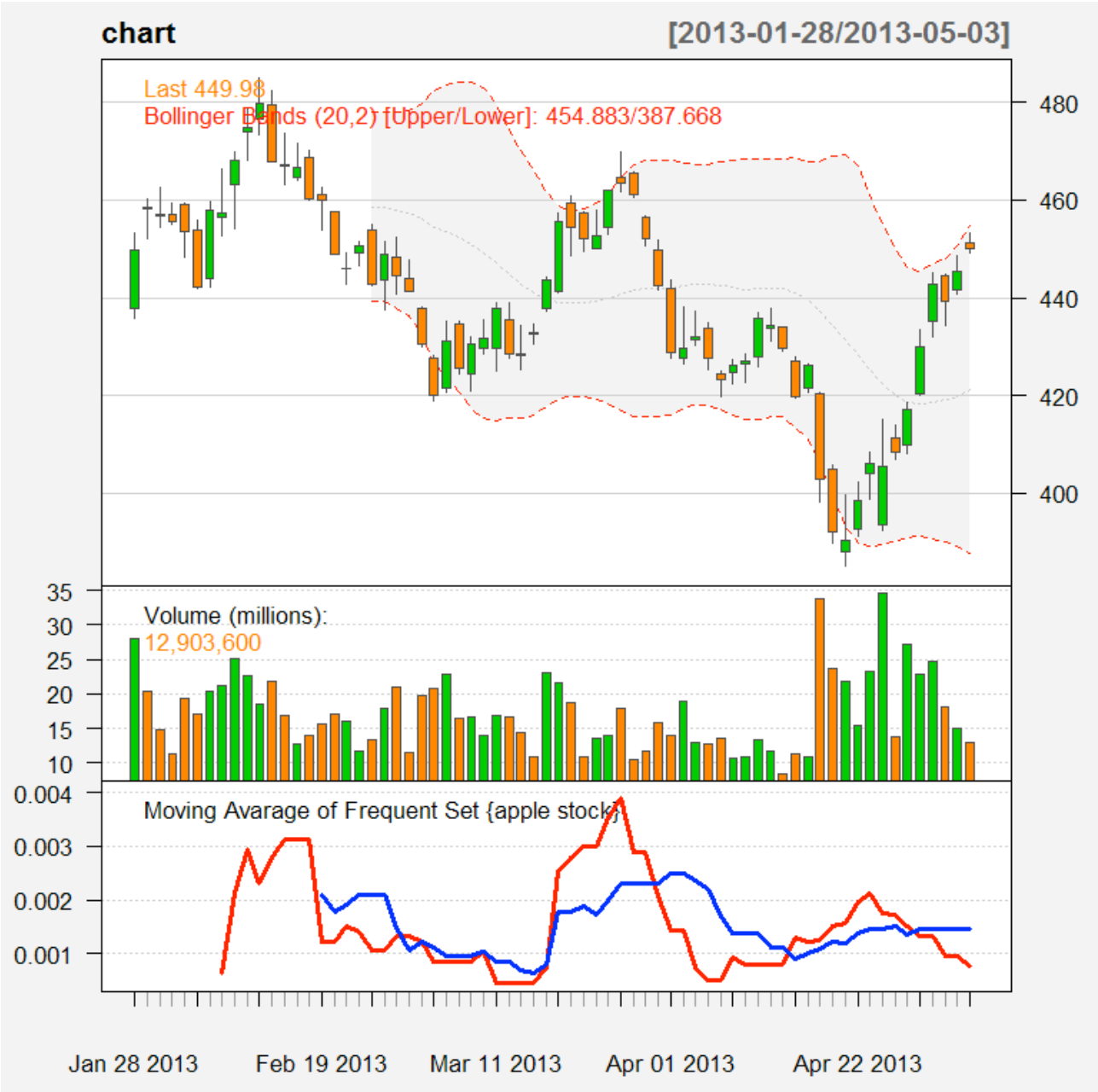

Fig.6 The stock chart with candles and volumes and moving averages of keywords frequent sets.

The developed package can also plot a crosscorrelation function, which shows the correlation between the moving averages of frequent sets and the stock price Fig.7). It allows to find predictive frequent sets.

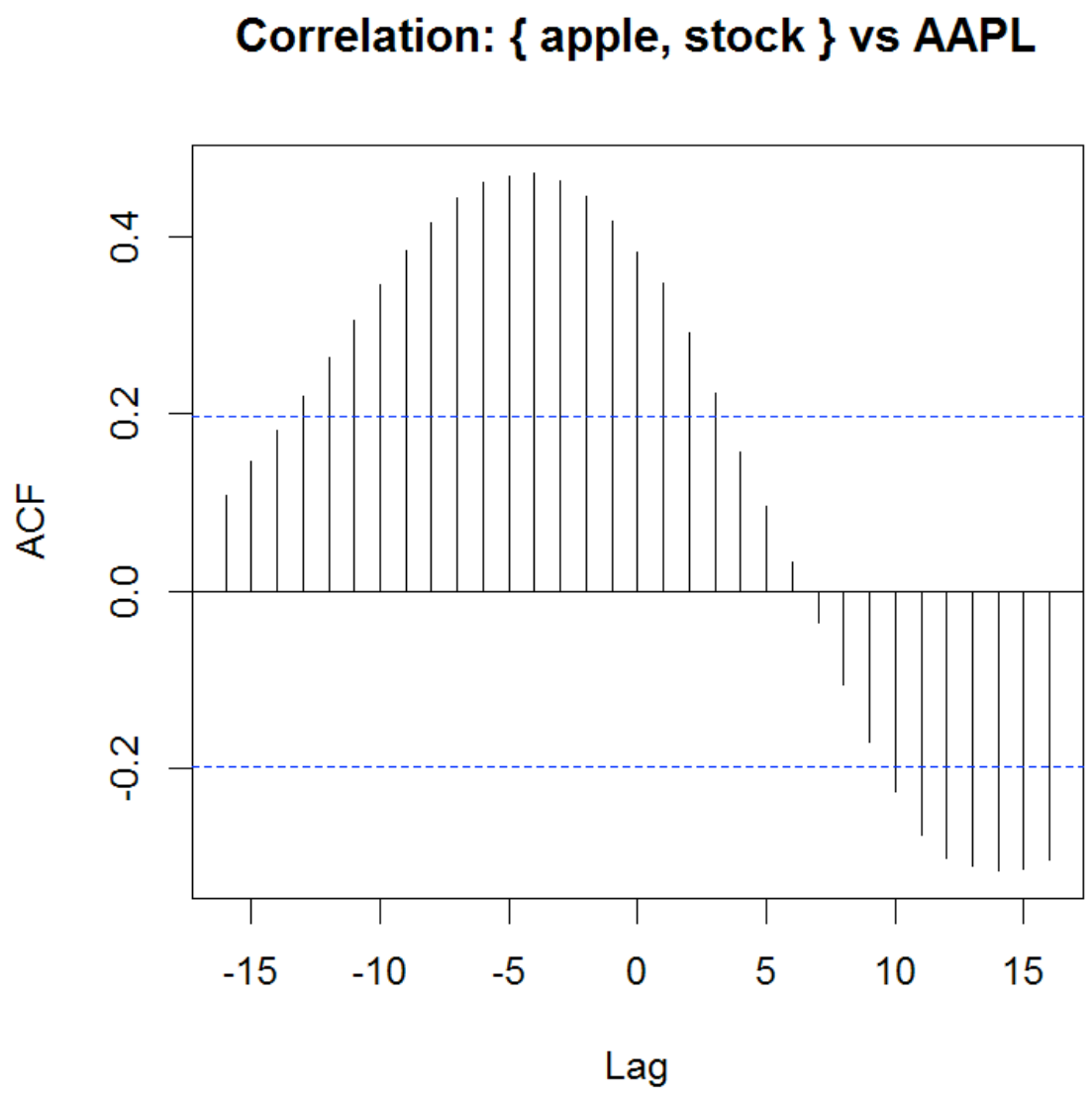

Fig.7 The crosscorellation between the stock chart AAPL and the moving averages of keywords frequent sets {apple,stock}.

This software also allows to perform the forecasting for the dynamics of stock charts; it is based on the ARIMA model using R package 'forecast' (Fig.8).

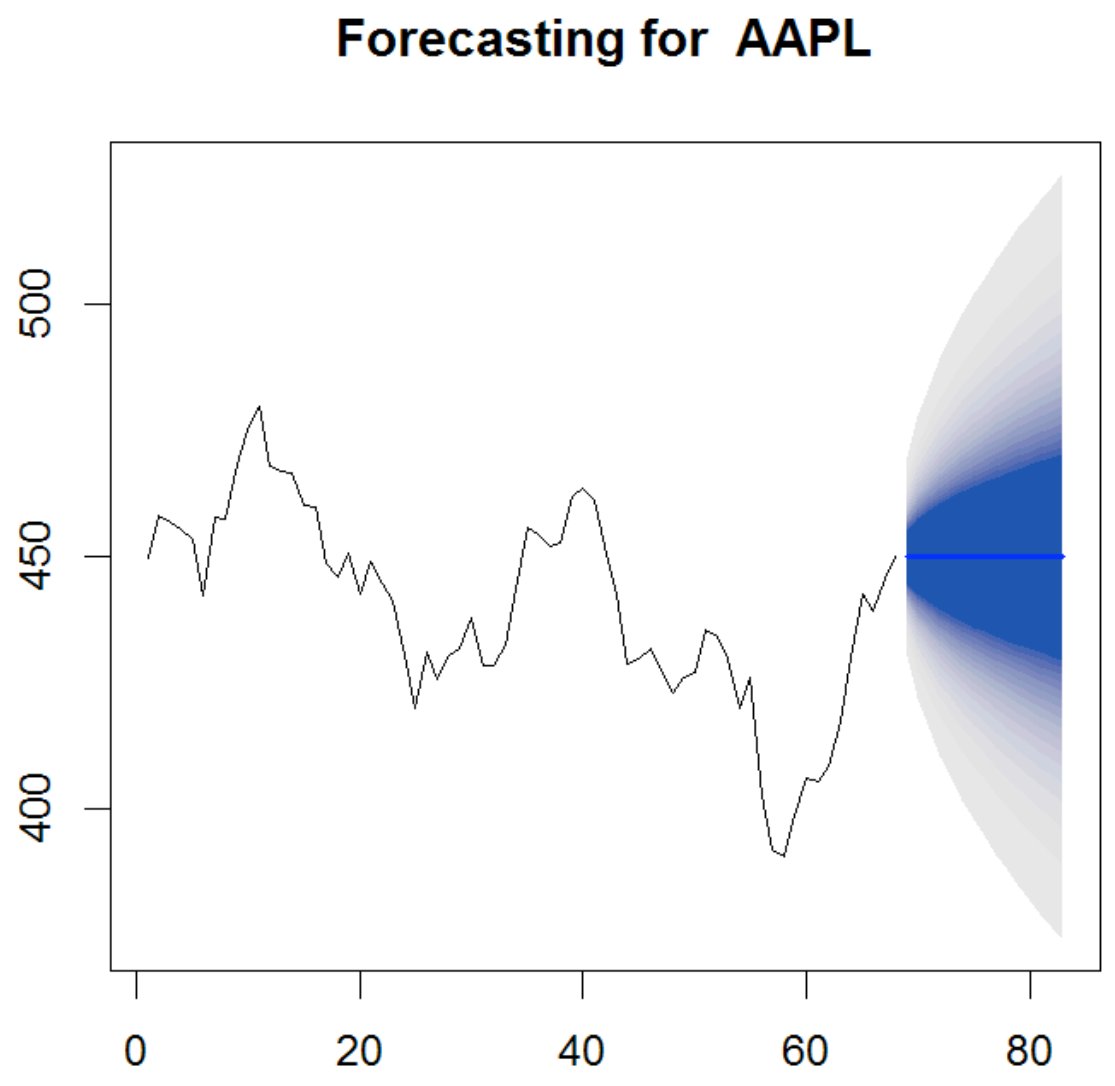

Fig.8 The forecasting of the stock chart AAPL using ARIMA model.

## Consideration and conclusions

Comparing the moving averages of some found frequent sets of keywords and the stock market, we can notice that these curves have similar trends regions, which are also proved by the calculated crosscorelation function. We have also performed an additional analysis, which is not included into the package under consideration. In the Fig.9, we showed the graph for the frequent itemsets, which is formed in the analyzed tweets array with some specified keywords. In the Fig.10 we showed the graph of association rules which are formed with the specified keywords. Such graphs can map the semantic structure of analyzed tweets array.

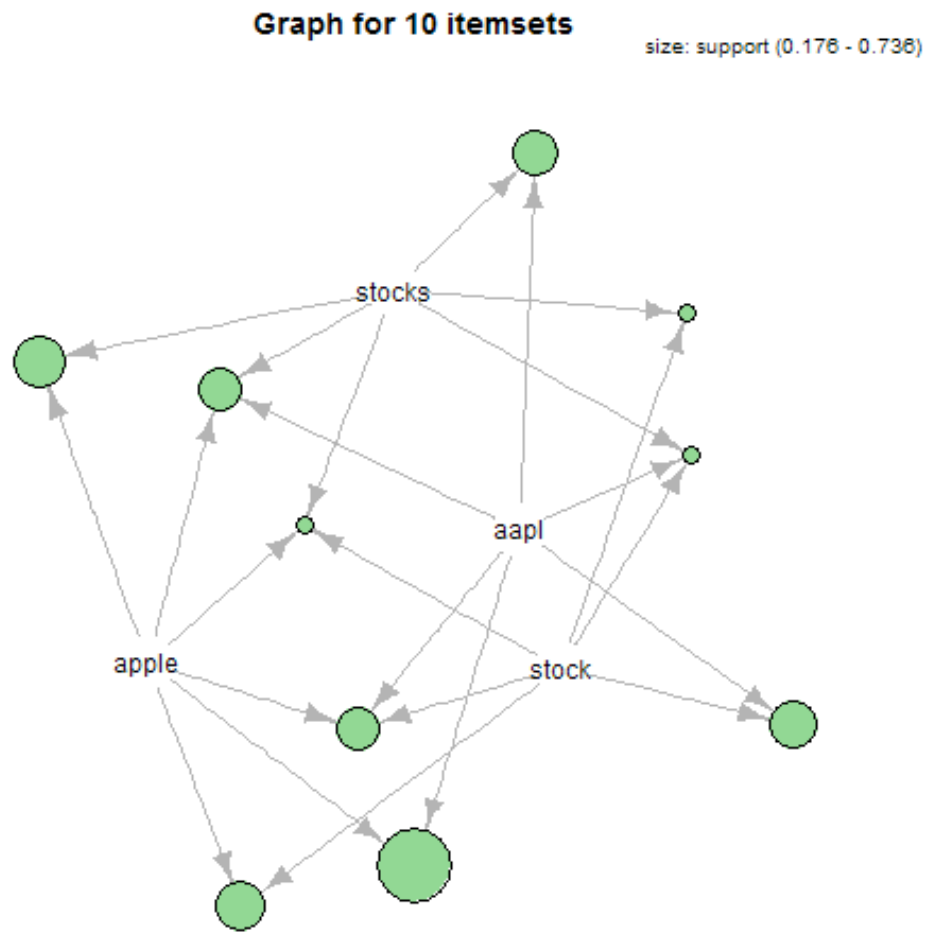

Fig.9 The graph for the frequent sets of specified keywords.

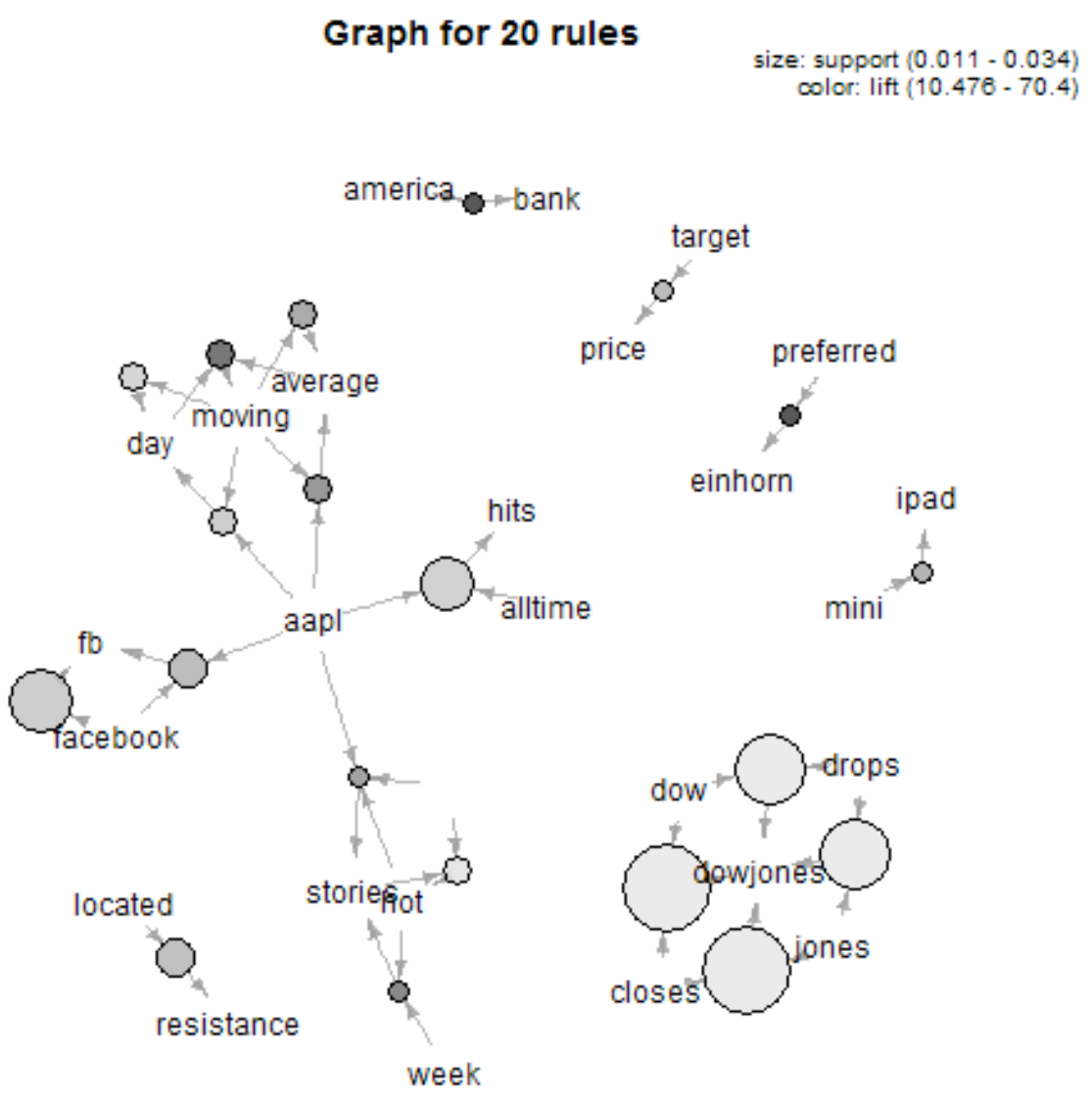

Fig.10 The graph for the association rules of specified keywords.

We conducted the Granger test to determine the causation between the time dynamics of frequent itemsets and the stock price. In the first test, we considered the null hypothesis about lack of causality between the dynamics of the frequent itemset {apple, stock} and AAPL stock price; in the second test, we examined the null hypothesis about lack of the causality between AAPL stock prices and the dynamics of the frequent itemset {apple, stock}. The calculations were performed using R packages. We have got the following results:

```
test 1
Granger causality test
Model 1: V3 ~ Lags(V3, 1:1) + Lags(V2, 1:1)
Model 2: V3 ~ Lags(V3, 1:1)
  Res.Df Df     F   Pr(>F)
1     87
2     88 -1 10.05 0.002103 **
---
Signif. codes:  0 '***' 0.001 '**' 0.01 '*' 0.05 '.' 0.1 ' ' 1

test 2
Granger causality test
Model 1: V2 ~ Lags(V2, 1:1) + Lags(V3, 1:1)
Model 2: V2 ~ Lags(V2, 1:1)
  Res.Df Df      F Pr(>F)
1     87
2     88 -1 0.3261 0.5694
```

p-value in the first test is equal to 0.002103, this is significantly less than the standard significance level of 0.05. The P-value in the second test is equal to 0.5694, this is substantially more than the standard significance level of 0.05. It means that the dynamics of the frequent itemsets of keywords {apple, stock} in users' tweets under analysis determines the dynamics of Apple stock prices.

Thus, the dynamics of support of some of the identified frequent itemsets of keywords reflects the trends of the stock market. Such frequent itemsets can be considered as potential predictive markers. Our next step is going to be the use of multivariate forecasting algorithms, based on the vector autoregressive model (VAR model). These algorithms can include many time series into analyses, the time series describe both stock prices and quantitative characteristics of tweets. We suppose such an approach will give the narrower and more precise forecasting. We are also planning to use the theory of semantic fields, frequent sets, association rules, Galois lattice, and the formal concepts analysis.